\documentclass[epj]{svjour}
\usepackage{graphics}
\begin{document}
\title{Violation of ensemble equivalence in the 
antiferromagnetic mean-field XY model}

\author{
Thierry Dauxois\inst{1}\thanks{Thierry.Dauxois@ens-lyon.fr},
Peter Holdsworth\inst{1}
\and
Stefano Ruffo\inst{1,2}\thanks{INFM and INFN, Firenze (Italy)}
}                     
%
%
\institute{Laboratoire de Physique, UMR-CNRS 5672,
ENS Lyon, 46 All\'{e}e d'Italie, 69364 Lyon C\'{e}dex 07, France
\and 
Dipartimento di Energetica "S. Stecco",
Universit\`a di Firenze, via S. Marta, 3, I-50139 Firenze, Italy}
\date{Received: date / Revised version: \today}
%
\abstract{ It is well known that long-range interactions pose serious
  problems for the formulation of statistical mechanics. We show in
  this paper that ensemble equivalence is violated in a simple
  mean-field model of $N$ fully coupled classical rotators with
  repulsive interaction (antiferromagnetic XY model). While in the
  {\it canonical} ensemble the rotators are randomly dispersed over
  all angles, in the {\it microcanonical} ensemble a bi-cluster of
  rotators separated by angle $\pi$, forms in the low energy limit. We
  attribute this behavior to the extreme degeneracy of the ground
  state: only one harmonic mode is present, together with $N-1$ zero
  modes. We obtain empirically an analytical formula for the
  probability density function for the angle made by the rotator,
  which compares extremely well with numerical data and should become
  exact in the zero energy limit. At low energy, in the presence of
  the bi-cluster, an extensive amount of energy is located in the
  single harmonic mode, with the result that the energy temperature
  relation is modified. Although still linear, $T = \alpha U$, it has
  the slope $\alpha \approx 1.3$, instead of the canonical value
  $\alpha =2$.
} 
%
\authorrunning{Dauxois, Holdsworth and Ruffo}
\titlerunning{Violation of ensemble equivalence in the 
antiferromagnetic mean-field XY model}
\maketitle

\section{Introduction}
\label{Introduction}

The relation between microscopic dynamics and macroscopic
thermodynamic behaviour can nowadays be explored in full detail in
computer simulations~\cite{Chaos}.  This allows one to test important
hypotheses of statistical mechanics.  Formost among these is
equivalence of ensembles.

It is widely accepted that the constant energy {\it microcanonical}
ensemble gives the same results for average values as the constant
temperature {\it canonical} ensemble.  This is true under certain
conditions~\cite{Rue}, among them the most important is that
interactions must be short-ranged. If interactions are long-range and
attractive, as for gravitating systems, all of thermodynamics breaks
down due to the non extensivity of thermodynamic
potentials~\cite{Pad90}. For systems of Coulomb charges of opposite
signs, screening effects may help in the construction of
thermodynamics, but the problem is not trivially solved, even for
classical systems~\cite{plasmas}, and each case must be examined
separately.

In this context, mean-field models occupy a special status.  Here,
thermodynamic potentials can be made extensive if the thermodynamic
limit is performed by rescaling the coupling with system size and
letting the range of the interaction go to infinity~\cite{Hem}.
Mean-field models are quite well studied in the {\it canonical}
ensemble and serve as a zero-order characterization of phase
transitions.  Although their solution is often trivial for systems
without disorder, they may hide important subtleties for disordered
and frustrated systems~\cite{Par}.  Far fewer studies of mean-field
models exist in the {\it microcanonical} ensemble. This may be the
reason why, until now, ensemble equivalence has not been questioned
for these models. In fact, one might have already expected some
surprises on the basis of the exact solution of a model by Hertel and
Thirring~\cite{Thi70}, where in the mean-field limit in the presence
of extensive thermodynamic potentials ensemble inequivalence was
explicitly shown.

In this paper we present a detailed study of the low
temperature/energy phase of a model of classical rotators, whose
potential energy is that of the mean-field antiferromagnetic XY model.
The ground state of this model is highly degenerate, and while in the
canonical ensemble equilibrium states are disordered at all
temperatures (i.e. rotators do not display any directional
organization), microcanonical ensemble simulations show the presence
of a ``bimodal" state where rotators are mainly grouped in two
``clusters", pointing in directions separated by an angle $\pi$.  This
dynamical effect also has thermodynamical consequences: the order
parameter measuring the degree of clustering is non-zero in the limit
of zero energy and the energy temperature relation is not that
predicted by the canonical ensemble.  We find that, although the
energy temperature relation $T=\alpha U$ is linear, as in the
canonical case, the coefficient is $\alpha \approx 1.3$ and not the
canonical value $\alpha=2$.  From the point of view of dynamics, the
rotators can be separated into two groups: a slow group which
oscillates around the bi-cluster and a group of almost freely rotating
rotators, which we call a ``gas'', following an analogy with particle
motion.

In Section~\ref{Model} we introduce the model. In Section~\ref{Ground}
we discuss the unusual properties of its ground state and in the
following Section~\ref{Statistical} we present the main controversial
points related to ensemble inequivalence. In Section~\ref{Structural}
we present an analytical expression for the probability density
function (PDF) for the orientation of a rotator, while the system is
in the bi-cluster state, and we explain the consequences for the
statistical properties in the system in the microcanonical ensemble.
In Section~\ref{Momentdy} we discuss the main features of the dynamics
of the moments of the PDF.  The paper ends with some conclusions and
perspectives.

\section{The model}
\label{Model}

We consider classical rotators denoted by the angle $\theta_i$, $i=1,\dots,N$,
which all interact  with each other, with an antiferromagnetic coupling $1/N$ and an
external field $h$
\begin{equation}
V=\frac{1}{2N} \sum_{i,j} \cos (\theta_i-\theta_j)- h \sum_i \sin \theta_i~.
\label{potential}
\end{equation}
We will mostly restrict ourselves to $h=0$.

After defining the complex order parameters (with $i$ the imaginary unit)
\begin{equation}
M_k= \frac{1}{N} \sum_n \exp (i k \theta_n)=|M_k| \exp (i \psi_k)~,
\label{order}
\end{equation}
the potential can be rewritten as
\begin{equation}
V= N \left( \frac{|M_1|^2}{2}  - h |M_1| \sin \psi_1 \right)
\label{newpot}
\end{equation}
The microcanonical ensemble is obtained by adding a kinetic energy term to the
above potential~\cite{Kogut,Antoni}
\begin{equation}
H=K+V~,
\end{equation}
with
\begin{equation}
K=\sum_{n=1}^N \frac{p_n^2}{2}~.
\end{equation}
In this formulation the model can also be thought of as describing a system
of particles with unitary mass, interacting through the mean-field
coupling $V$ (the names rotator and particle will be equivalently used
throughout the paper). This model has been first introduced in Ref.~\cite{Kogut}
for 2D nearest-neighbour couplings and then studied in the antiferromagnetic
mean-field context in Ref.~\cite{Antoni}.

The total energy $E= U N$ ($U$ being the energy density) is fixed by
the initial conditions and is conserved in time, while temperature is
defined through the time averaged kinetic energy (see Ref.~\cite{temp}
for alternative definitions of temperature), $T=2<K>/N$, where
$<\cdot>=\displaystyle\lim_{t \to \infty}\displaystyle {1\over t}
\int_0^t$.  If $h=0$, as is usually the case, then the total momentum
$P=\displaystyle \sum_n p_n$ is also conserved. We set $P=0$ in order
to avoid ballistic center of mass motion when $h=0$.

The equations of motion
\begin{equation}
\ddot{\theta}_n= |M_1| \left[ \sin (\theta_n - \psi_1) + h \cos \theta_n \right]
\end{equation}
have been integrated using an improved fourth-order symplectic scheme~\cite{Mac}.
The algorithm is ${\cal O}(N)$, provided one first computes $M_1$ in the central loop.
During the time evolution, we sample the instantaneous values of the order parameters
(\ref{order}) up to $k=20$ and we compute their running time averages.
System size was varied from $N=100$ to $N=10^4$.
We have also performed canonical Monte-Carlo simulations, using the Metropolis algorithm,
for comparison.

The initial conditions were of two classes: {\it i)} homogeneous state
$\theta_n=(2 \pi n)/N$, which can be shown to be marginally stable
(see below), to which we add either a small spatially random
perturbation $\theta_n \to \theta_n+ r_n$ and/or a small momentum
$p_n=r_n$ with zero average; {\it ii)} homogeneous state $\theta_n =
(2 \pi n)/N$ with $p_n = A \sin \theta_n$, which leads to a faster
induction of the bimodal state we want to study. These initial
conditions lead to the same bimodal state, discussed below, at low
energy.

The canonical solution of this model for $h=0$ is sketched in Ref.~\cite{Antoni}. It uses
the Hubbard-Stratonovich trick to decouple the rotators and the thermodynamic limit
is performed by a saddle-point technique. The result is that all moments (\ref{order})
vanish in the $N \to \infty$ limit, including the magnetization $M_1$, which
implies, on the one hand, that
\begin{equation}
T=2U~,
\label{equi}
\end{equation}
on the other, that since in this limit
\begin{equation}
M_k=\int_0^{2\pi} {\cal P}(\theta) \exp (ik\theta) d\theta~,
\end{equation}
the PDF ${\cal P} (\theta)$ is flat. Hence, the bimodal state is absent in the canonical
ensemble, which we have confirmed by Monte-Carlo simulations.

Most of the discussions below will concentrate on the reasons for the different
findings in the microcanonical and canonical statistical ensembles.
 
\section{Ground State and Statistical Properties in the Canonical Ensemble}
\label{Ground}

The long ranged interactions mean that it is impossible to satisfy all 
the antiferromagnetic bonds at once and the model is highly frustrated.
In zero field, the frustration is minimized for configurations with 
$M_1 = 0$, giving a ground state energy of $U=E/N=0$.

The ground state is infinitely degenerate, as there is an 
infinity of ways of minimizing the frustration. For example, grouping
the rotators into pairs with angles $\theta_i$ and $\theta_i +\pi$
ensures that $M_1=0$ for all configurations of the pairs. The
pairs do not have to be arranged in an ordered way however, as 
any disordered arrangement will equally give 
$M_1=0$. Neither is the ground state manifold restricted to pairs: one
can equally construct ground states from groups of three, four, five ...
rotators separated by angles of $2\pi/3, 2\pi/4, 2\pi/5...$. By moving
rotators in clusters whose total angle is zero the system can evolve from 
one ground state to another remaining on the constant energy hypersurface.

The high dimensional ground state manifold follows from the fact that
the ground state condition requires the two constraints $M_{1x} =M_{1y}=0$
only.
One can therefore expect a ground state to possess $N-2$ unconstrained
degrees of freedom, which is easily confirmed by calculating
the Hessian
$J_{i,j}=- \partial^2 V/\partial \theta_i \partial \theta_j$. For example, 
for the perfectly homogeneous ground state at $h=0$, with $\theta_i=(2 \pi i)/N$
\begin{equation}
J_{i,j}= - \frac{1}{2N} \cos \left( \frac{2 \pi (i-j)}{N} \right)~.
\label{Hess}
\end{equation}
The matrix indeed has two non-zero equal eigenvalues, $-1/4$ and $N-2$
zero eigenvalues.

Collective organization of the particles into reduced symmetry states
can reduce the number of constraints even further.  For example, in
our system, if the rotational symmetry is broken and the spins lie
along a single axis the ground state condition is reduced to a single
constraint $M_x=0$, leading, at the harmonic level to $N-1$ free
degrees of freedom~\cite{Moessner}. This result is confirmed by
calculating the Hessian (\ref{Hess}), on a perfect bi-cluster
groundstate with $N/2$ rotators at $\theta = 0$ and $N/2$ at
$\theta=\pi$. One finds only one non-zero eigenvalue, $-1/2$,
corresponding to the counter vibrating motion on the circle of the two
groups of particles; all the other eigenvalues are zero.

The model we study is an extreme case of a collective
paramagnet~\cite{Villain}, or classical spin
liquid~\cite{Moessner,CHS}; a system that remains disordered with no
evidence of spin freezing down to the limit of zero temperature.  The
special points on the ground state manifold with a reduced number of
constraints can dominate the partition function, in the canonical
ensemble, leading to an ``Order by Disorder''
transition~\cite{Villain2} to a reduced symmetry state. However, the
mode counting arguments of Moessner and Chalker~\cite{Moessner}
predict this to happen only if the number of liberated modes at the
special points exceeds the number of zero-modes in a state with full
symmetry. This is certainly not the case here and as we have confirmed
by Monte Carlo simulation, Order by Disorder does not occur.  We have
simulated between $N=10$ and $N=10^4$ rotators down to 
temperatures $T < 10^{-4}$. At all temperatures, the system remains perfectly
disordered, with no evidence of bi-cluster formation.  The number of
zero modes can be directly verified by measuring the specific heat at
constant field, $C_h$, 
at low temperature, as each quadratic mode makes a contribution
$1/2$ (in units of $k_B$), while each zero mode makes a contribution
zero. We find $C_h = 1.0$, for the $N$ rotator system, as expected for
two regular modes. There is therefore no evidence of the system
preferring states with a single quadratic mode.

\section{Statistical Properties in the Microcanonical Ensemble}
\label{Statistical}

The system reserves a surprise, when studied in the microcanonical
ensemble, as we do not observe ensemble equivalence. For the classes
of initial conditions cited above, the low energy state of the system
is not one with a homogeneous distribution of angles, rather we
observe the formation of a bimodal structure, with enhanced
probability for two angles separated by a distance
$\pi$~\cite{Antoni}. As the energy goes to zero, the asymptotically
reached state has broken symmetry, but is not perfectly bi-modal, as
we discuss below.  Nevertheless, when the Hessian for such a state
is calculated numerically, we find that it also has only one non-zero
eigenvalue. It seems therefore that the formation of the bicluster in
the microcanonical ensemble, realises the condition of minimizing the
number of non-zero modes.

\begin{figure}
\null
\vskip 1truecm
\resizebox{0.6\textwidth}{!}{  \includegraphics{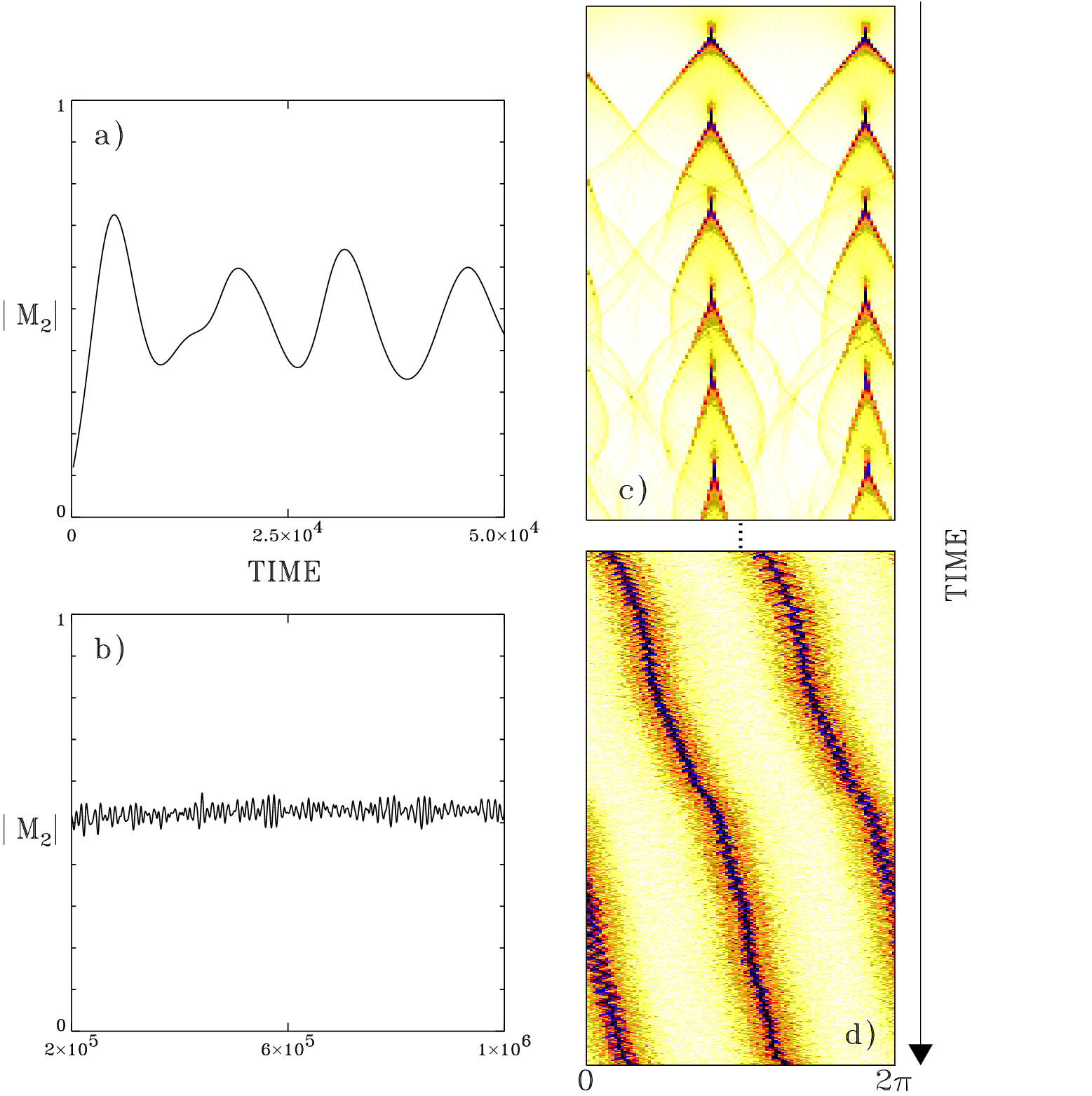}}
\vskip 1truecm
\caption{(a) (resp. (b)) shows the instantaneous value of $|M_2|$ as
 a function of time
at early (resp. later) times. (c) (resp. (d)) shows the evolution in
grey scales of the distribution ${\cal P} (\theta,t)$ at early
(resp. later) times. The $\theta$ axis is horizontal, the time axis
vertical pointing downwards.}
\label{dynamics}
\end{figure}

After a transient time $\tau$, the formation of a stable bi-cluster is
revealed by a non-zero value of the second moment $|M_2|$ of ${\cal P}
(\theta)$. The early time evolution of $|M_2|$ shows the typical
behavior observed for the growth of the mean-field in self-consistent
dynamical models~\cite{Tenny} with an initial exponential growth,
followed by a saturation reached after a damped oscillatory motion has
died away, see Figs.~\ref{dynamics}(a) and \ref{dynamics}(b).  In
Figs.~\ref{dynamics}(c) and \ref{dynamics}(d) we show the space-time
plot of ${\cal P} (\theta,t)$, which reveals the origin of the
oscillations in the density waves originated periodically from the
bi-cluster, preceding stabilization.  The darkest colours correspond
to the highest densities. A very sharply peaked but unstable
bi-cluster forms over a short time period. The structure disperses
with well defined instability edges that propagate from the cluster
center and re-appears in a quasi-periodic manner with slowly
lengthening period. The dispersing cluster appears to interact with
propagating fronts from previous incarnations. The result is that the
dispersion is successively slower and the fronts less well defined for
following quasi-periods until finally the bi-cluster stabilizes. In
addition Fig.~\ref{dynamics}(d) shows that once this coherent
structures has emerged and is stabilized, it propagates around the
circle with apparently ballistic dynamics.

\begin{figure}
\resizebox{0.5\textwidth}{!}{  \includegraphics{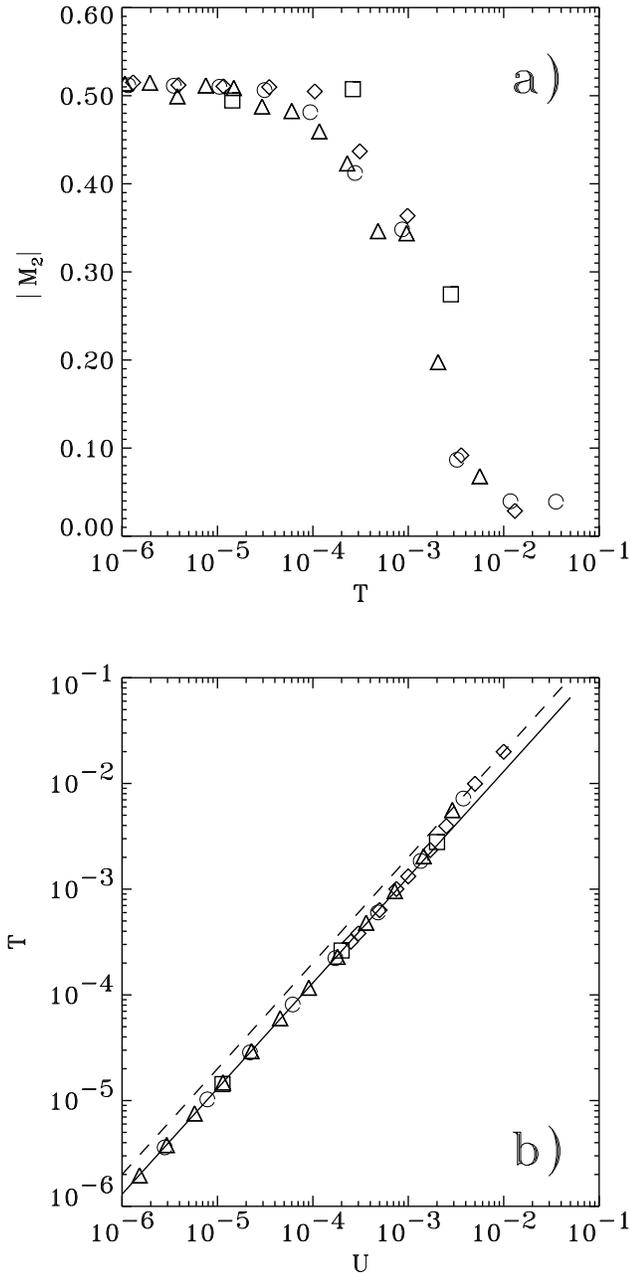}}
\caption{(a) $|M_2|$ as a function of the temperature $T$ 
  for $N=200$ (triangles), $N=500$ (circles), $N=10^3$
  (diamonds), $N=10^4$ (squares).  (b) $T$ vs. $U$ for the same values
  of $N$.  The solid line is the modified microcanonical relation $T =
  1.3\ U$, whereas the dashed one corresponds to the canonical
  relation $T=2\ U$ and is valid in the high energy regime, where the
  cluster has disappeared ($|M_2|=0$).}
\label{um2t}
\end{figure}

In Fig.~\ref{um2t}(a), we show $|M_2|$, averaged over times later than
$\tau$, as a function of temperature $T$, for various system sizes.
For low temperatures ($T<10^{-4}$), $|M_2|$ approaches the value
$|M_2| \sim 0.5$ and never decays, even in extremely long simulations
(times as long as $10^8$ in proper units\footnote{The appropriate
linear time scale of the system is $2 \pi={\cal O}(1)$}). On the contrary, in the
high temperature regime ($T>10^{-2}$) the bi-cluster never forms,
$|M_2| \sim 0$ for all times and the PDF remains flat, as in Monte-Carlo
simulations. In the intermediate regime, $ 10^{-4}< T < 10^{-2}$, the
bi-cluster forms but is less well-defined, corresponding
to progressively smaller values of $|M_2|$. As the temperature
is increased there is a smooth transition to the homogeneous state
observed in the canonical ensemble.  In the next Section we will
further discuss the internal structure of the bi-cluster.

The choice of initial conditions discussed above is dictated by the
need to progressively increase the energy starting from a ground
state. One may wonder what happens for more general initial
conditions. 
We therefore looked at statistics  of the final configuration, starting from many different realizations of a uniformly random
distribution of angles and zero momenta. For $N=100$, at temperatures 
around $T \sim 10^{-5}-10^{-4}$, the bimodal state was always reached
over $500$ initial states.  For $N=1000$ and $T \sim 10^{-6}$,
the totality of 250 random initial states went to the
bimodal state.  It means that the bi-cluster is fully attractive for
this class of initial states, which is quite generic.

As the formation of the bi-cluster is a unique property of the
deterministic microcanonical system, it is an example of violation of
the equivalence hypothesis between microcanonical and canonical
behaviour. One might therefore expect this inequivalence to show up in
other measurable quantities. This is indeed the case, for example we
find that the standard relationship between energy and temperature is
modified.  In the canonical ensemble, as there are only two harmonic
modes, equipartition of energy at low temperature leads to the energy
temperature relation $E=NT/2+T$ ($N$ quadratic modes for the kinetic
energy and 2 for the potential energy), which implies that $T=2U+{\cal
  O}(N^{-1})$, see (\ref{equi}). The eventual Order by Disorder 
averaging over parts of
the phase space with only one quadratic modes (or any finite number of
them) produces only $1/N$ corrections to this relation. In the
microcanonical simulation, in the presence of the bi-cluster we find
the anomalous relation $T \sim 1.3U$ as shown in Fig.~\ref{um2t}(b).
Once the temperature exceeds $T>10^{-2}$, the bi-cluster does not form
and the energy-temperature relation crosses over to the canonical
expression. In the presence of the bi-cluster, the kinetic energy is
therefore much smaller than one should expect and the mean potential
energy far in excess of that predicted by equipartition for the single
non-zero mode associated with the bicluster.  In fact, from this
result, we find that the configurational contribution to the heat
capacity $C_h \approx <V>/T \sim (0.35 N)/1.3$ is an extensive
quantity, in complete contradiction to the predictions of mode
counting in the canonical ensemble.

The system manages to put a macroscopic amount of energy in the single
non-zero mode corresponding to the vibrating motion of the two groups
of rotators and gives the system a low-dimensional aspect with the
bi-cluster taking on many characteristics of a two-particle system.
Fluctuations in the magnetization $|M_1|$, whose mean value is
identically zero in the ground state illustrate this point.  In zero
field the potential energy is $V = N|M_1|^2/2$, hence the extensive
nature of the configurational heat capacity implies that mean value
of  the magnetization $<|M_1|>$ should be of order unity and not
of order $1/\sqrt{N}$ as one might expect in an uncorrelated
paramagnetic system.  We have indeed observed ordered features in the
time dependence of $M_1$ which are consistent with the
non-fluctuational value of this quantity.

\section{Structural and Dynamical Details of the Bi-Cluster}
\label{Structural}

\begin{figure}
\resizebox{0.45\textwidth}{!}{  \includegraphics{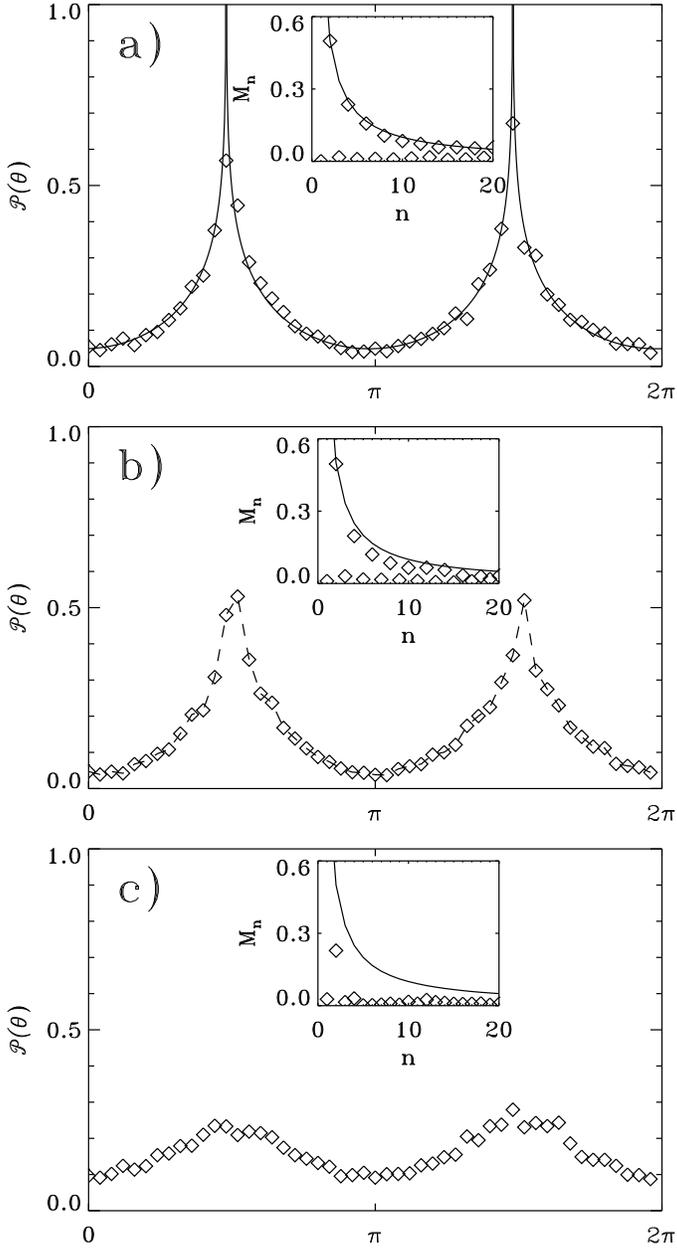}}
\vskip 1truecm
\caption{${\cal P} (\theta)$ for $N=10^4$ and three different values of 
$T$: $1.46 \cdot 10^{-5}$ (a), $2.6 \cdot 10^{-4}$ (b), $2.8 \cdot 10^{-3}$ (c).  
The full line in (a) is the analytical formula
(\ref{distr}), the diamonds are the numerical histograms (in 
(b) we join them for convenience with a dashed line).  In the insets we show the
corresponding momenta of the PDF (diamonds), the full line is $1/n$.
In (a) the moments respect the $1/n$ law for even $n$, while in (c)
only the mode $n=2$ is non-zero. Odd modes are much smaller.}
\label{histogram}
\end{figure}

The bi-cluster never becomes perfectly formed, with $|M_2| = 1$, even in 
the limit $U \to 0$. Rather, the angular distribution of rotators retains a width, with
a certain population of rotators homogeneously distributed. In Fig.~\ref{histogram}
we show the angular PDF ${\cal P}(\theta)$ for 
a system of $N=10^4$ particles and three different temperatures: one very low 
(Fig.~\ref{histogram}(a)) and two in the transition region 
(Fig.~\ref{histogram}b) and (c)). 
The distribution is peaked for two angles, separated
by $\pi$, but there remains non-zero probability of finding angles between
the two peaks.
The degradation of the bicluster is progressive as the temperature is increased.
We have made the experimental observation that the moments of the PDF assume the
following values as the energy is decreased
\begin{eqnarray}
<|M_k|>&\approx&0 \; \mbox{for odd k} \cr
<|M_k|>&=&1/|k| \; \mbox{for even k with} \, <|M_0|>=1~,
\label{moments}
\end{eqnarray}
as shown in the insets of Fig.~\ref{histogram} for the first $k=20$ modes. This observation
offers no contradiction to our finding that $<|M_1|>$ is independent of system size. As the
bi-cluster is stable at low temperatures only, the numerical value of $<|M_1|>$ remains small,
even though it is an intensive quantity. By summing
the Fourier series for ${\cal P}(\theta)$,
\begin{equation}
{\cal P}(\theta) = \sum_{k=-\infty}^{\infty} <|M_k|> \exp (-i k\theta)~,
\end{equation}   
one gets~\cite{Grad}
\begin{equation}
{\cal P}(\theta) = \frac{1}{2\pi} \left( 1 - \log (2 |\sin \theta| ) \right)~.
\label{distr}
\end{equation}   
This analytical formula is superimposed on to the numerical data in Fig.~\ref{histogram}(a)
with no free parameter apart from a shift of $\theta$, due to
the motion of the bi-cluster (see below). The agreement is impressive, and although
we have no theory for this result, we may well say we have a solution.
We expect the analytical formula (\ref{distr}) to become exact in the $U \to 0$
limit.

\begin{figure}
\resizebox{0.5\textwidth}{!}{  \includegraphics{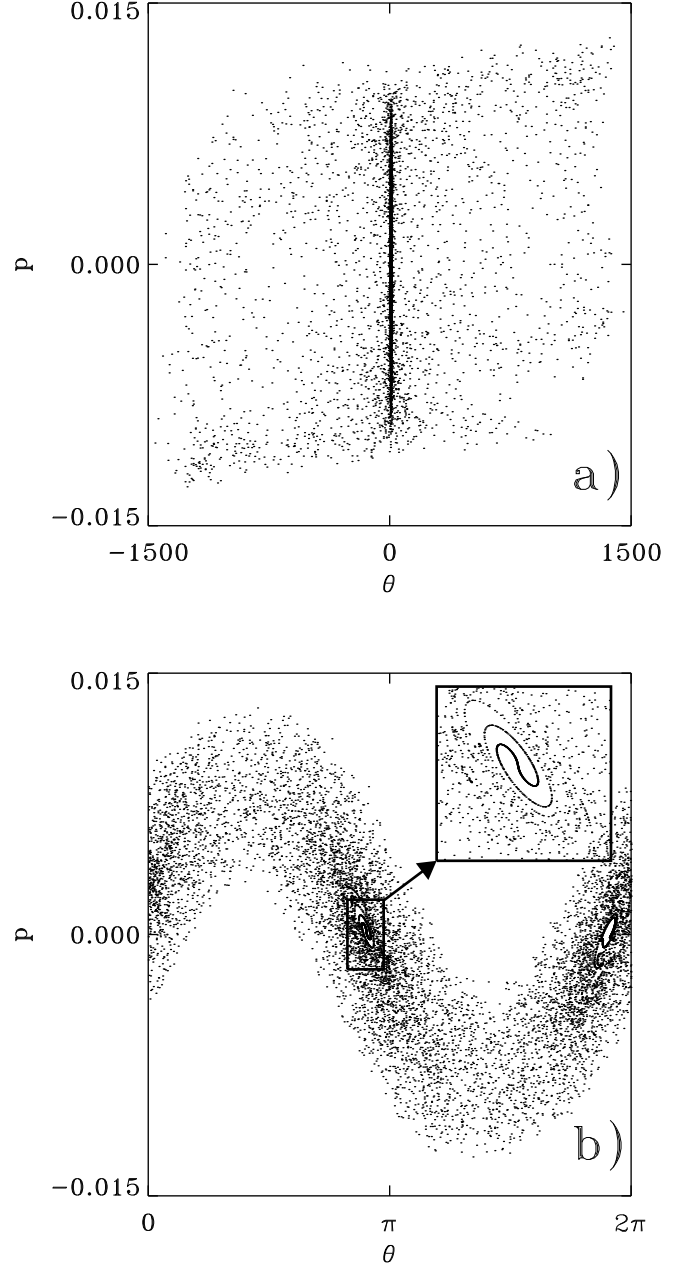}}
\caption{Single particle phase-space for $N=10^4$, $U=1.13 \cdot 10^{-5}$
and $t=2.5 \cdot 10^{5}$. In (b) the angle $\theta$ is folded onto $[0,2\pi[$
and the inset emphasizes the structure of the spirals arm around one of the bi-cluster center. }
\label{thetaimp}
\end{figure}

The internal structure of the bi-cluster is further shown in
Fig.~\ref{thetaimp}(a), where the state at time $t=2.5 \cdot 10^5$,
with $N=10^4$ and $T=1.43 \cdot 10^{-5}$, is displayed in the
$(\theta,p)$ single-particle phase-space (so-called Boltzmann
$\mu$-space). A large fraction of particles get stuck in the
bi-cluster and have small values of $|\theta|$, while others
develop much larger $\theta$ values. This illustrates why we speak
of a ``gas-cluster'' coexistence: particles appear to belong to
two distinct groups; those in the bi-cluster, which perform
oscillations around the two centers separated by angle $\pi$, and those 
in the gas, which have ballistic dynamics and travel over large
distances. Once projected onto the $p$-axis the distribution is
symmetric (giving, for instance, zero average momentum), but as we see
it in the $(\theta,p)$ space it is skewed, with, on average, bigger
momenta for those particles which have traveled the furthest.

\begin{figure}
\resizebox{0.5\textwidth}{!}{  \includegraphics{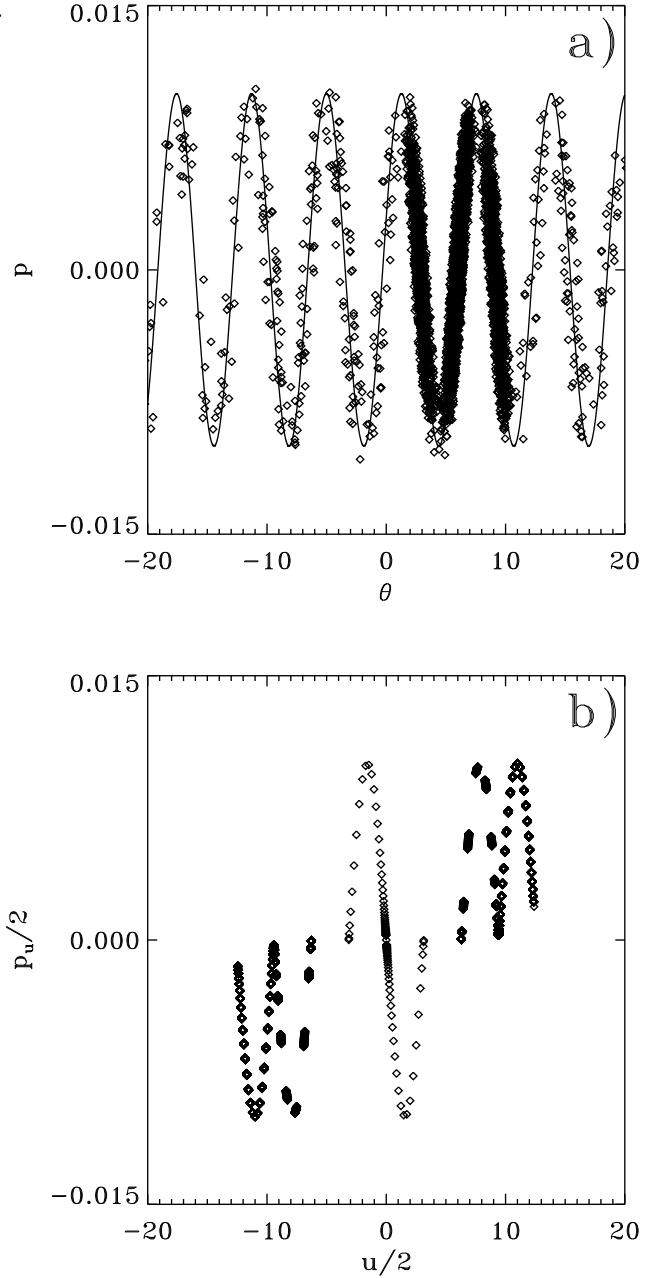}}
\caption{(a) Zoom of the central region of Fig.~\ref{thetaimp}a, showing the phenomenological
law $p=A \sin (\theta + \phi)$ (\ref{sin}) with $A=0.01$ and $\phi=0.3$. (b) Phase-space of
$10^4$ initial particles in the square $(u,p_u)=[-\pi*0.0005,\pi*0.0005]^2$
of the toy model (\ref{mod}) (with $\Delta t=1$) at iteration $3000$. 
}
\label{serpent}
\end{figure}

This picture is further clarified if we fold the $\theta$-axis onto $[0, 2\pi]$
(Fig.~\ref{thetaimp}b). The oscillating motion of the particles of the cluster is now
revealed by the spiral arms visible in the bi-cluster centers (see the inset). The particles in the cluster appear now to be distributed 
in a sinusoidal band of lesser density. 
Indeed, if we look inside the region around $\theta=0$
(Fig.~\ref{serpent}a), we see that the particles are quite well distributed along the
phase line 
\begin{equation}
p=A \sin (\theta+\phi)~,
\label{sin}
\end{equation} where $\phi$ is a sliding phase related to the
bi-cluster motion and $A$ an energy related amplitude. This gives us a hint 
as to how to discover the dynamical mechanism at work in the process of particle 
evaporation from the bi-cluster. In fact, differentiating (\ref{sin}) with respect 
to time and using $\dot{\theta}=p$, one gets for $u=2\theta$ the equation
of motion
\begin{equation}
\ddot{u}= A^2 \sin u~,
\label{invpend}
\end{equation}
which describes a pendulum with gravity pointing upwards, $u=0$ being a saddle-point.
This system is integrable and cannot produce the phase-distribution in Fig.~\ref{serpent}(a).
We must, therefore introduce some sort of perturbation capable of producing the sinusoidal
layer observed in Fig.~\ref{thetaimp}(b). A well known mechanism for producing a stochastic
layer in pendulum motion is the introduction of a finite time step, as done
in the Chirikov standard map. We have therefore decided to iterate the map
\begin{eqnarray}
u'&=& u + \Delta t\  p_u \cr
p'_u&=& p_u + A^2 \Delta t\  u'~,
\label{mod}
\end{eqnarray}
with $\Delta t=1$, taking an ensemble of initial points homogeneously
distributed in a small square around the unstable point $u=p_u=0$. A
snapshot of these points at iteration $3000$ is shown in
Fig.~\ref{serpent}(b); the particles have been stretched along the
separatrix layer and distributed along it in a similar way to that shown in 
Fig.~\ref{serpent}(a).  Two features are responsible for the stretching:
the unstable manifold distributes the particles away from the saddles,
while the stable manifold attracts them to the saddles.  The
inhomogeneous distribution of the points is a result of these two
mechanisms.  One feature of the evaporation in the full model, which is
not at all reproduced by the map (\ref{mod}), is the fact that time
evolution seems to select only one of the two lobes of the separatrix;
in other words, the symmetry $p \to -p$ seems to be broken in the full
model, while stretching in both the lobes is present in the map
(\ref{mod}).

This map is of course unable to explain why the bi-cluster forms and
why it is stable.  That is, if the evaporation mechanism were the only
process present then the bi-cluster would be progressively depleted.
This is in contrast with numerical simulations for low temperature
values, which show the temporal stability of this collective state.
Hence, a condensation mechanism on the bi-cluster should also exist,
in order to establish a steady-state. Indeed, Fig.~\ref{thetaimp}(b)
shows that in the single-particle phase space two elliptic points are
present at the centers of the bi-cluster, therefore in the bordering
chaotic region of these two points a ``trapping'' effect could be
present.

If formula (\ref{sin}) is approximately verified, then the knowledge
of the PDF at low temperatures (\ref{distr}) would allow us to compute
the relationship between the temperature and the constant $A$.
Performing the integral, it turns out that
\begin{equation}
T=<p^2>=\frac{A^2}{4}~.
\label{ptoA}
\end{equation}
This relation is very well verified numerically for low temperature
values. Moreover, particle momenta appear to follow a distribution
which, although non-Gaussian, has a variance given by (\ref{ptoA}).
The distribution shows a sharp peak in the center, due to the core of
the bi-cluster, which is far from Gaussian. However, the wings of the
distribution, which are due to the ``gas'' do follow a Gaussian
distribution.

One might propose that total momentum conservation plays a role in the
observed phenomenology.  An easy way to test this idea is by adding a
small external magnetic field $h$, which removes the constraint. We
observe that, for small $h$, the bi-cluster still forms and is stable
at low temperatures. The conservation law is not
therefore relevant for cluster formation.  On increasing $h$, the
distribution ${\cal P}(\theta)$ is modified, as shown in Fig. 6. The
bi-cluster lies along the field axis and becomes asymmetric, with the
number of rotators lying parallel and antiparallel to the field
direction becoming unequal.  The antiparallel cluster is continuously
depleted until a single cluster eventually forms. This route towards
a single cluster is somewhat counter intuitive if one thinks of a
N\'eel ordered, unfrustrated antiferromagnet. Such a system would
minimize its energy by aligning the bi-cluster perpendicular to the
field and relaxing the two halves continuously out of the plane in a
symmetric way. The bi-cluster would be destroyed by the two clusters,
of equal size, rotating continuously onto the field direction.

\begin{figure}
\resizebox{0.5\textwidth}{!}{  \includegraphics{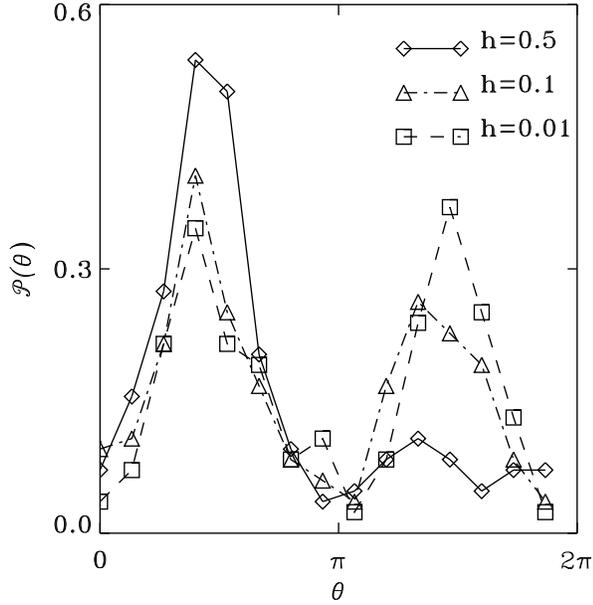}}
\caption{${\cal P}$($\theta$) in the presence of an external field for a 
chain of $N=200$ particles. The diamonds correspond to $h=0.5$ whereas the triangles to $h=0.1$
and the squares to $h=0.01$.}
\label{field}
\end{figure}

\section{Moment dynamics}
\label{Momentdy}

As we have already remarked, there is no implicit contradiction
between the proposed formula for the moments of the PDF
(\ref{moments}) and the presence of a modified energy temperature
relation. The latter means that $<|M_1^2|>$ is an intensive quantity
rather than being ${\cal O} (1/N)$ as in the in the canonical
ensemble.  Hence, although all the other odd modes of the PDF vanish
in the $N \to \infty$ limit, this is not true for $<|M_1|>$. Rather,
it remains finite in the low temperature regime, increasing linearly
$U$. Solution (\ref{moments}) is therefore exact, only when $U \to 0$.
However, because of the smallness of $U$ in this regime, $<|M_1|>$ is
a small quantity with respect to $<|M_2|>$ and equation
(\ref{moments}) is perfectly valid.

\begin{figure}
\resizebox{0.5\textwidth}{!}{%
  \includegraphics{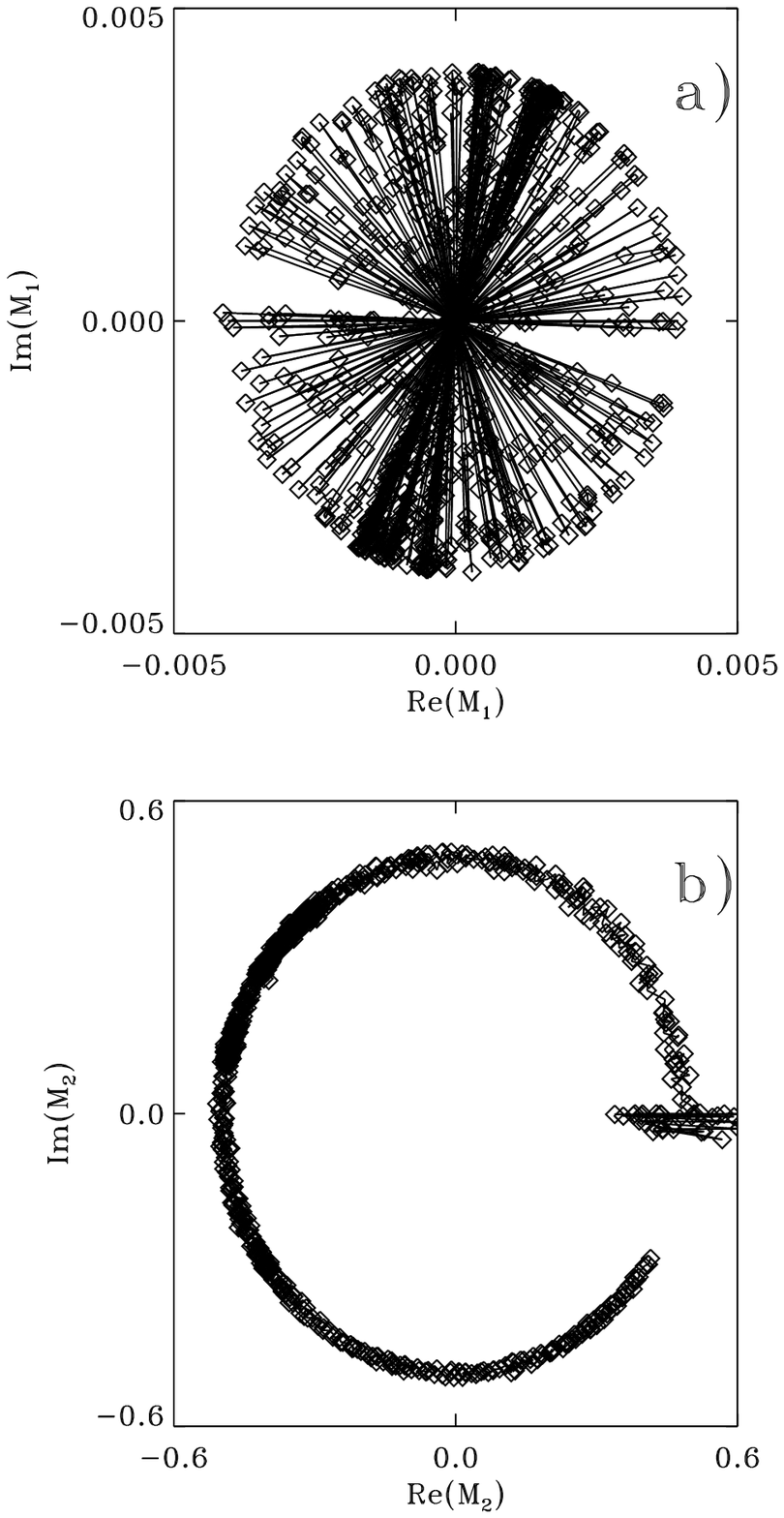}}
\caption{(a) Phase-points of $M_1$ for $N=10^4$ particles and 
$U=1.13 \cdot 10^{-5}$.
(b) Same for $M_2$.}
\label{mm2af}
\end{figure}

We expect that $M_1$ will display interesting dynamical behaviour,
given its rather unexpected extensive nature, when the bi-cluster is
formed.  The $($Re$(M_{1})$,Im$(M_{1}))$ phase-plane is shown in
Fig.~\ref{mm2af}(a) for $N=10^4$ and $U=1.13 \cdot 10^{-5}(T=1.43
\cdot 10^{-5})$. Successive points are joined by lines to show the
relevant properties of the motion. The phase-point has a fast
oscillatory motion through zero and a much slower rotatory motion
centered on zero. The fast motion is due to the vibrations of the
bi-cluster around the equilibrium positions (the two components of
$M_1$ cross zero in phase). The slow rotational motion is due to the
rigid rotation of the bi-cluster.  This latter motion is further
revealed by the dynamics of $M_2$ in Fig.~\ref{mm2af}(b).  Again we
join successive phase-points with a line, showing that the phase point
in the $M_2$ plane is rotating around the center, maintaining a fixed
radius $|M_2| \approx 0.5$.  The time dependence of the phase
$\psi_2$ of $M_2$ is such that, over the long time span, the average
is zero, but preliminary measurements of the variance $\sigma^2$ show
that the motion is ballistic, rather than diffusive, $\sigma^2 \sim
t^2$.

This picture is not only qualitative, but also quantitative. Indeed,
if we average the fast oscillatory motion of $|M_1|$ in
Fig.~\ref{mm2af}(a), we get an estimate of $<|M_1|^2>$. Taking
$<|M_1|^2>/U \approx 0.7$, we get the correction to the energy
temperature relation $T \approx (2 - 0.7) U = 1.3 U$ as shown in
Fig.~\ref{um2t}(b).

At the transition temperature where $|M_2|$ is decreasing, both the
motion of $M_1$ and that of $M_2$ become more erratic, revealing the
beginning of the region where the bi-cluster is progressively
depleting in time.

The other odd moments of the PDF always show an erratic motion around
zero, with the variance of the cloud of points decreasing as $N$ is
increased; i.e. higher odd modes are not intensive. The higher even
moments of the PDF show a pattern similar to that of
Fig.~\ref{mm2af}(b), with a progressively reduced radius; the motion
of the higher phases has also ballistic features.

\section{Conclusions and perspectives}
\label{Conclusions}

The antiferromagnetic mean-field classical rotator system is an ideal
laboratory to study the relation between microcanonical and canonical
ensembles. Although it has a trivial canonical solution, the randomly
uniform state at all temperatures, the high degeneracy of its ground
state induces nontrivial dynamical effects, which are revealed in the
microcanonical ensemble. Instead of maintaining a random distribution
of the rotators, the Hamiltonian dynamics selects a bimodal state,
where the rotators are oriented along angles at distance $\pi$ with
some spread, a bi-cluster. We have introduced an order parameter which
reveals the formation of this state in the low energy phase and we
have empirically obtained an analytical formula for the probability
distribution function in angle, which perfectly fits the numerical
data.

In addition to this first remarkable difference between the two ensembles, we
have also shown that the energy temperature relation is modified and,
although still linear, it has a different slope in the two
ensembles. The origin of this behavior lies in the extensive amount of
energy which Hamiltonian dynamics puts into the oscillatory vibrating
motion of the bi-cluster.

Mean-field models are characterized by self-consistency. Indeed, a
feature of our model is that rotators generate themselves the
mean-field in which they move. Therefore, in the large $N$ limit, the
dynamics of our model should depend only on the interaction of the
single rotator with the mean-field.  As already claimed for other
models (e.g. beam-plasma instability and vorticity defect
model~\cite{Tenny}) self-consistency effectively reduces the number of
degrees of freedom. This is why many properties of the dynamics of
such a complex $N$-body system as ours resemble those of a ``simple''
perturbed pendulum. This is also the origin of the ordering of the
rotators in a bi-cluster. There should be entropic reasons why the rotators
prefer this state rather than choosing the disordered state, which is
instead selected in the canonical ensemble.

Many questions remain to be explored, but the most
compelling one concerns the careful description and explanation of the
dynamics. The perturbed pendulum analogy must be further investigated
and a thorough study of the time evolution of the moments of the
probability distribution function in angle should allow a better
understanding of the low-dimensional properties of the dynamics.

\acknowledgement We thank M. Droz, J. Farago, M-C Firpo, M. Paliy and
Z. Racz for useful discussions.  S.R. thanks ENS-Lyon, INFN and INFM
for financial support. P.H. thanks INFN for financial support. This
work was performed using the computing resources of the P\^ole
Scientifique de Mod\'elisation Num\'erique (PSMN) of ENS Lyon and of
the DOCS-INFM group in Florence.

\end{document}